# A MULTI MEGAWATT CYCLOTRON COMPLEX TO SEARCH FOR CP VIOLATION IN THE NEUTRINO SECTOR

L. Calabretta, M. Maggiore, L. A. C. Piazza, D. Rifuggiato, INFN-LNS, Catania, Italy
A. Calanna, CSFN-SM, Catania, Italy

*Abstract*

A Multi Megawatt Cyclotron complex able to accelerate $H_2^+$ to 800 MeV/amu is under study. It consists of an injector cyclotron able to accelerate the injected beam up to 50 MeV/n and of a booster ring made of 8 magnetic sectors and 8 RF cavities. The magnetic field and the forces on the superconducting coils are evaluated using the 3-D code OPERA. The injection and extraction trajectories are evaluated using the well tested codes developed by the MSU group in the '80s. The advantages to accelerate $H_2^+$ are described and preliminary evaluations on the feasibility and expected problems to build the injector cyclotron and the ring booster are here presented.

## INTRODUCTION

Recently members of the neutrino community proposed a new experiment called DAEδALUS (Decay At rest Experiment for $δ_{cp}$ At Laboratory for Underground Science) a new approach to search for CP violation in the neutrino sector [1,2]. They proposed to utilize high power proton accelerators able to supply a proton beam with about 800 MeV, 1.5 MW average power and a duty cycle of 20% (100 msec beam on, 400 msec beam off). DAEδALUS needs three sources of neutrino; the nearest one located at 1.5 km from the underground detector must have a minimum power of 1 MW. The second source should stay at a distance of about 8 km from the detector and should supply an average beam power of 2 MW or more. The last neutrino source, 20 km far from the detector, has to be fed with a proton beam of average power higher than 5 MW. The neutrino fluxes produced by the three sources are measured by the 300 KTons Cerenkov detector filled with water doped with Gd.

The three sources have their beam time synchronised, so the detector will receive the 100 msec beam bunch from each source in sequence, while for a 100 msec all the three sources will be kept off, to allow the measure of the background.

This configuration allows to measure, how many oscillations $\bar{\nu}_\mu \rightarrow \bar{\nu}_e$ occur for each source.

Although the required average power for the first 2 sites is 1-2 MW, the 20% duty cycle has the consequence that the peak power is 5-10 MW and a peak current of about 9.5 mA is necessary. At the same time the lower beam power mitigates the problems related to thermal dissipation and activation.

At a current higher than 1 mA the space charge effects become more and more relevant both for the injection process and for the extraction process. Solutions that mitigates the spaced charge effects are advantageous, see next section.

Accelerator complexes consisting of two or three cyclotrons, one or more injector cyclotrons and a main ring cyclotron booster, have already been proposed as drivers for energy amplifier or waste transmutation plants [3,4,5]. The main constraints for these accelerator complexes are: current higher than 10 mA and energy as high as 1 GeV, minimum beam losses, high reliability and high conversion efficiency from electrical to beam power.

We believe that up to now accelerator driven systems (ADS) based on well known conventional cyclotrons accelerators are the most reliable and economical solution for a plant which requires a peak beam power of 1-5 MW[4,5]. To deliver higher peak power, i.e. 10 or more MW, the key points for the ring cyclotrons are the space charge effects, the extraction devices and the power to be dissipated in each cavity. To overcome these problems a classical solution is to increase the radius of the cyclotron and the number of cavities. But this means to increase significantly the plant cost.

An alternative solution based on the acceleration of $H_2^+$ molecule has been proposed [6,7]. In this case the extraction of the $H_2^+$ beam is accomplished by a stripper which produces two free protons breaking the molecule. Due to the different magnetic rigidity as compared to the $H_2^+$, the protons escape quite easily from the magnetic field of the cyclotron. Extraction by stripping does not require well separated turns at the extraction radius and allows using lower energy gain per turn during the acceleration process and/or lower radius for the magnetic sectors, with a significant reduction of thermal power losses for the RF cavities and construction cost. The extraction by stripper allows to extract beams with large energy spread (0.5÷1%) so the energy spread produced by space charge effect on the longitudinal size of the beam is not crucial in this kind of accelerator, and flattopping cavities are unnecessary.

We believe that the acceleration of $H_2^+$ beam, despite it needs to handle beam with magnetic rigidity two times larger, offers a remarkable advantage in terms of reliability, easier operations and lower cost.

In the past, a layout for an accelerator complex able to supply a proton beam with energy of 1 GeV and a beam power up to 10 MW was presented by some of the authors [6], in the perspective to drive a sub critical reactor. This previous proposal is now updated to fit the requirement of the MIT scientists. Moreover the number of accelerators required by the experiments, at least 4-5, forces us to minimize the accelerator cost.

The solution, here presented, consists in a two cascade cyclotrons complex. The injector cyclotron, a four sector machine, accelerates a beam of $H_2^+$ up to energy of about 50 MeV/n. The beam is then extracted by an electrostatic deflector and it is injected from the injector cyclotron into an 8 sectors Superconducting Cyclotron Ring. Two stripper foils are used to extract two proton beams at the same time from the ring cyclotron. This solution allows increasing the mean life of the stripper foils and reducing greatly the design of the beam dump.

## SPACE CHARGE EFFECTS

Before the injector cyclotron description, we like to underline the problems related to the space charge effects for $H_2^+$ vs. proton beam. The space charge produces a repulsive force inside the beam bunches, which generate detuning effects. To evaluate the strength of this effect the parameter called "generalized perveance" has been introduced [8]. The generalized perveance is defined by the following formula:

$$K \propto \frac{qI}{m \cdot \gamma^3 \beta^3}$$

Where: $q$, $I$, $m$, $\gamma$ and $\beta$ are respectively the charge, current, mass and the relativistic parameters of the particle beam. From this formula it is quite evident that the proton beam has a perveance double as compared to the $H_2^+$ beam when the two particles have the same speed and the same current. But if protons and $H_2^+$ are accelerated by the same electric field, they have the same energy but not the same speed. On the other hand a beam of $H_2^+$ delivers a number of protons which is double as compared to a proton beam with the same current.

In the first 3 rows of Table 1, we compare the perveance of $H_2^+$ beam and proton beam with a current of 5 mA and 10 mA respectively, at various energies.

The ratio of perveance values shows that, concerning the space charge effects, to accelerate a $H_2^+$ beam is more convenient than a proton beam with a double current. This advantage is also more evident when energy increases. The last two rows of Table 1 show the perveance values of a proton beam with a current of 2 mA and the ratio vs. the perveance of $H_2^+$ beam with 5 mA. Although the perveance of the 2 mA proton beam is lower at low energy, we see that if the energy of the $H_2^+$ beam is increased of a factor 2.3 the same perveance value of the proton beam is achieved.

Table 1: Perveance values of proton and $H_2^+$ beams at various energies.

|  | $E_p=E_{H2}$ 30 keV | $E_p=E_{H2}$ 800 MeV | $E_p=30$ keV $E_{H2}=70$ keV |
|---|---|---|---|
| $H_2^+$, I=5 mA | 0.881 10$^{-3}$ | 0.151 10$^{-9}$ | 0.247 10$^{-3}$ |
| P, I=10 mA | 1.245 10$^{-3}$ | 1.075 10$^{-9}$ | 1.245 10$^{-3}$ |
| $K_{H2}/K_p$ | 0.707 | 0.141 | 0.198 |
| P, I=2 mA | 2.491 10$^{-4}$ | 2.15 10$^{-10}$ | 2.491 10$^{-4}$ |
| $K_{H2}/K_p$ | 3.537 | 0.703 | 0.992 |

## INJECTOR CYCLOTRON

The injector II of the PSI and the commercial compact cyclotron designed by EBCO and IBA companies are the only kind of cyclotron accelerators which are able to deliver more than 1.5 mA of proton beam up to now.

The injector II of PSI is a conservative solution which is able to supply up to 3 mA of proton beam. It is a separate sector cyclotron with beam injection at 800 keV, final energy of 70 MeV, energy gain per turn ≅1 MeV, extraction radius of 3.3 m and single turn extraction by electrostatic deflector. Despite low voltage injection (25-30 keV) and moderate energy gain per turn (<200 keV/turn), the compact commercial cyclotrons are able to accelerate proton beams with current of 1.5-2.2 mA [9], but these accelerators use the stripper extraction to deliver the beam.

According to the above evaluation, the perveance of a $H_2^+$ beam with a current of 5 mA and with energy of 70 keV is similar to the perveance of a proton beam with 2 mA and energy of 30 keV.

For the previous reasons, we propose a design which is a mixing of the PSI injector II and of the compact commercial cyclotron described before. The central region of the proposed injector is a scaled up central region of the commercial cyclotron. To take account of the higher magnetic rigidity and to maintain the perveance of the $H_2^+$ beam similar to the perveance of the proton beam injected into a commercial cyclotron, both the injection energy and the energy gain per turn are doubled. Moreover, the energy gain per turn is supposed to increase along the radius up to the value of 1.8 MeV at the extraction radius. This value is higher than the energy gain per turn in the PSI injector to compensate for the smaller extraction radius. Although the final turn separation at the extraction is 12 mm in the solution here proposed, while for the PSI injector II the turn separation is about twice, we believe that beam losses should be lower than 0.2%. This means that expected beam losses at the extraction should be lower than 400 W for a delivered average beam power of 200 kW. Lower beam losses could be achieved increasing the voltage of the cavities at extraction. Moreover the transverse and longitudinal emittances of the beam injected into the ring cyclotron can be larger than for the PSI ring cyclotron, indeed in our solution extraction is performed by strippers which do not need a single turn extraction to achieve extraction efficiency of 99.99%.

Table 2: Parameters of the Injector Cyclotron

| Einj | 35 keV/n | Emax | 50 MeV/n |
|---|---|---|---|
| Rinj | 41.6 mm | Rext | 1.440 mm |
| <B> at Rinj | 1.29 T | <B> at Rext | 1.39 T |
| Sectors N. | 4 | Cavities N. | 4 |
| RF | 30 MHz | harmonic | 3$^{rd}$ |
| V at Rinj | >70 kV | V at Rext | 250 kV |
| Injection eff. | 15 % | Extraction eff. | 99.8% |
| ΔR at Rext | 11.6 mm | ΔE/turn | 1.8 MeV |
| Δx at Rext | <3.5 mm | Turns N. | < 83 |
| Extraction: Electrostatic Deflector + Magnetic Channels | | | |
| Deflector Gap | 12 mm | Electric field | <50 kV/cm |

The tentative parameters for the injector cyclotron are presented in Table 2. The beam injection through the central region is a poor efficiency process, about 15% without buncher and may be 20-25 % with buncher, so a lot of injected beam will be lost along the first turn. We estimated that along the first 3 "posts", which are used to select the proper longitudinal phase acceptance, the beam power lost will be of about 3kW at each post. Although the low energy of the beam does not produce activation anyway, the short range of the particles and the high power lost will pose serious problem of thermal cooling.

## $H_2^+$ Ion Source

To accelerate a beam current with a peak current of 5 mA, due to the low efficiency injection, we need a source of $H_2^+$ able to deliver a beam current of 24 or 35 mA respectively if a buncher is used or not. Despite parasitic beam of $H_2^+$ is ever produced in any kind of proton source, up to now ion source able to deliver the request intensity of $H_2^+$ are not yet reported. Anyway at LNS in Catania [10] a compact ECR Versatile Ion Source (VIS) able to deliver up to 32 mA of proton beam has been recently developed. Tests of VIS show a parasitic beam of $H_2^+$ that reaches up to the 20% of the proton beam. According to the designer of VIS an optimisation of the source parameters, like position of the permanent magnets, vacuum pressure, RF power, could allow to achieve a beam current of $H_2^+$ higher than 20 mA. Other two important parameters of VIS are its good beam normalised emittance, about 0.1 π mm.mrad, and its extraction voltage which could be raised up to 70 kV. Both these two parameters fit with the request of the injector cyclotron. Another important feature of VIS is the moderate construction costs.

An alternative ion source much more performing that VIS is the ion source under construction for the IFMIF project [11]. This source is designed to supply Deuteron beam current higher than 100 mA at 100 keV with a normalized emittance of about 0.3 π mm.mrad.

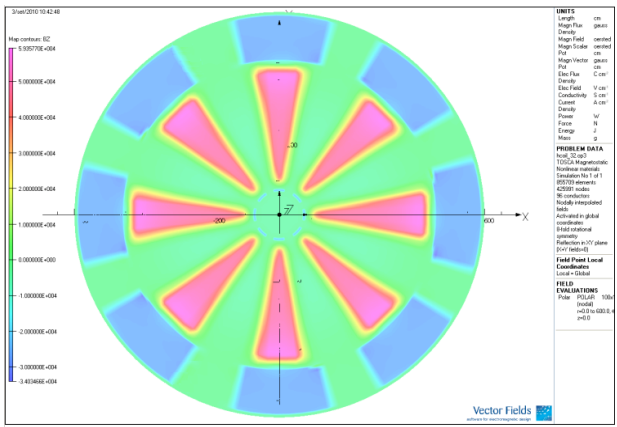

Figure 1: Magnetic field map of the ring cyclotron

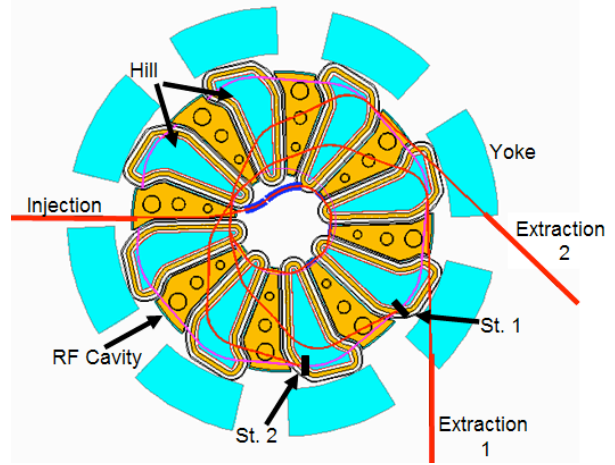

Figure 2: Layout of the ring cyclotron

Despite this is a source for deuterons the preliminary test to optimize the source optic will be performed by an $H_2^+$ beam, which is very similar to the deuteron beam. So this ion source prototype could be used to supply the $H_2^+$ beam for the injector cyclotron. Unfortunately the cost of this source is more expensive than the VIS cost.

## RING CYCLOTRON

One of main constraints on the accelerator design for the DAEδALUS experiment is to reduce the construction cost. For this reason our first attempt was to design a Ring Cyclotron with extraction radius of about 4 m. In Table 3 the main parameters of this preliminary study for the Ring Superconducting Cyclotron are presented. Our preliminary study was made using straight sectors. The pattern of the achieved magnetic field is shown in Fig. 1. The simulated configuration was able to produce a magnetic field near the required isochronous one with maximum differences lower than 5%, but the vertical focusing was not acceptable for radius higher than 3.4 m. To achieve a good vertical focusing it is mandatory to spiral the magnetic field up to achieve a spiral angle near 40° at the extraction radius.

Table 3: Main parameter of the RSC

| | | | |
|---|---|---|---|
| $E_{max}$ | 800 MeV/n | $E_{inj}$ | 50 MeV/n |
| $R_{ext}$ | 4.05 m | $R_{inj}$ | 1.44 m |
| <B> at $R_{ext.}$ | 2.28 T | <B> at $R_{in}$ | 1.39 T |
| Bmax | <6.3 T | Pole gap | >50 mm |
| $\xi_{spiral}$ | < 40° | Hill width | 20° |
| Outer radius | ≤6 m | Flutter | 1.7÷1.27 |
| Sector height | < 5 m | N. Sectors | 8 |
| Sector weight | < 300 tons | N. Cavities | 8 |
| Cavities λ/2 | Double gap | harmonic | 6th |
| RF | 59 MHz | V | 250-300 kV |
| ΔE/turn | 3.2 MeV | RF Power | 200 kW |
| ΔR at $R_{ext}$ | 1.5 mm | ΔR at $R_{inj}$ | > 15 mm |
| Coil size | 20 x 40 cm² | Icoil | 4300 A/cm² |

The magnetic field with spiralled sectors was used to evaluate both the injection and extraction trajectories, see Fig. 2. The beam envelopes along these trajectories and along the last equilibrium orbit at 800 MeV/n were also evaluated and are presented in Fig. 3. The normalised beam emittance in the transverse planes, used in the simulation, is 0.4 π mm.mrad.

The extraction by stripper was simulated simply changing the charge to mass ratio of the particles. There is a broad range of azimuth positions where it is possible to place the stripper to achieve a trajectory escaping the cyclotron field. This range starts just before the exit of the hill and ends just at beginning of the following hill. The trajectories starting in the middle of the valley pass very close to the centre of the cyclotron while the trajectories near the hill boundaries have a larger distance from the centre. These last trajectories do not cross the injection trajectory and stay inside the vacuum chamber. Like shown in Fig. 2, it is possible to use two strippers placed 45° apart one from the other to obtain two extraction trajectories. This is a good advantage because it allows doubling the mean life of the stripper foils and at the same time simplifies the design of the beam dump. Two or more beam dumps are acceptable for DAEδALUS experiment. Although the simulations are made with a preliminary field map, the results show that injection and extraction processes are feasible. To simulate the beam envelope and the trajectories we used the computer codes developed at MSU, with some minor changes.

Unfortunately the injection trajectory simulation was limited by the integration code, which uses the azimuth angle as independent variable. So this code does not allow simulating trajectories which bend clockwise. This constraint has strongly restricted the evaluation of other more convenient injection trajectories.

However the results of our simulations demonstrate that it is possible to inject and extract the beam without interference among the stripper extractions and the injection trajectories, moreover the vertical beam size along the trajectories which stay inside the cyclotron pole are smaller than 1 cm. The beam envelope of the extraction trajectories have a significant radial blow up just near the region where the beam escapes from the cyclotron field.

It is evident that to build a coil whit spiral shape is not very easy, moreover this solution reduces significantly the room for the RF cavities, which should have also a spiral shape. Due to the small room between the sectors and the spiralled shape it is difficult to insert a single gap, pill box cavity, like the PSI one. Moreover the use of the PSI like resonators needs more room at the inner radii where the present configuration is quite full by the injection line. A preliminary RF cavity, type λ/2, double gap, which fits in the empty space between the magnetic sectors was studied and its layout is shown in Fig.4. Although the RF cavity is able to produce enough high voltage, a satisfactory energy gain per turn with a moderate power loss of 200 kW, its shape poses serious problems for the installation. Indeed the radial insertion of the cavities is not feasible. The cavities have to be inserted from the top of the ring.

A further serious problem is given by the magnetic forces on the coils. Both the hoop stress on the coils and the resulting radial shifting force are too high. For these reasons we plan to update the present design increasing both the injection energy and the extraction radius. The preliminary results of the new simulation, with extraction radius of 4.5 m, give a flutter of 2.3, measured along a circle trajectory. The true flutter value when evaluated along the beam trajectory increases up to 3.5. This difference is due to the large scalloping of the trajectories which at the extraction radius reach up to 10 cm. The value of $v_z$ is now about 1 up to energy of 780 MeV/amu with completely straight sector. Work is in progress to achieve the proper isochronous field and to minimize the hoop stress on the coils and mainly to minimise the radial shifting force. Unfortunately, the maximum magnetic field on the coil in the new configuration is around 6.4 T. This high magnetic field exceeds the usual limit to use NbTi superconducting cable. We have to evaluate if it is more convenient to use Nb3Sn superconducting cable or to reduce the maximum magnetic field down to values below 6 T.

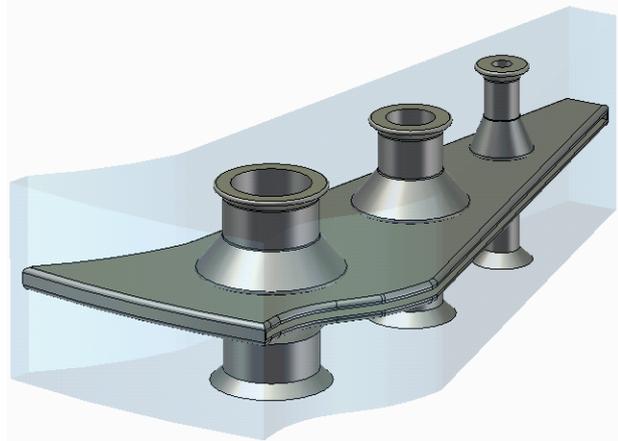

Figure 4: Layout of electrodes and stems of the proposed RF cavity. The three stems are useful for mechanical stability and to host the cryopanels to pump inside the vacuum chamber.

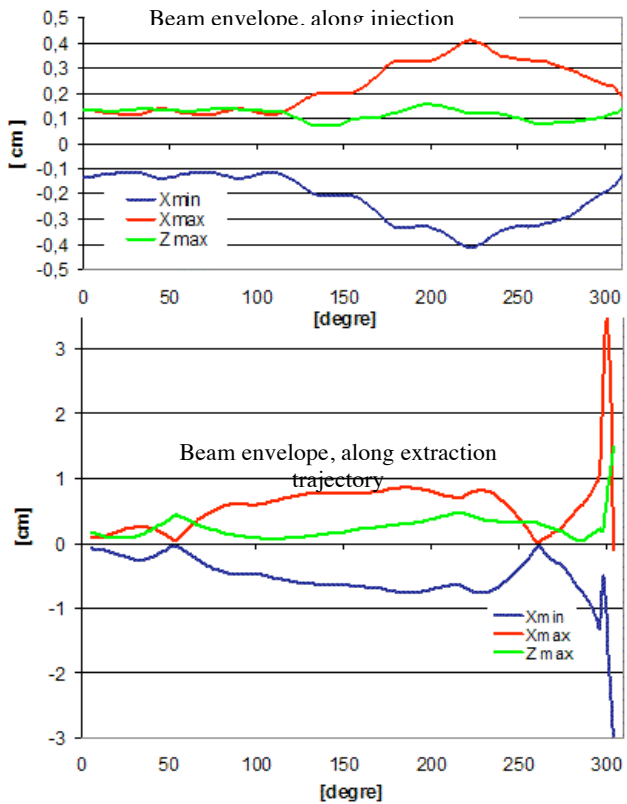

Figure 3: Beam envelope along the injection trajectory, upper, and along the extraction trajectory, lower. The 0° is the matching point on the equilibrium orbit at the injection or at the extraction radii; the 300° is outside the cyclotron.

## EXTRACTION BY STRIPPING

The stripping extraction is largely used in many commercial cyclotrons to deliver intense proton beam with energy of 30 MeV. This consist to accelerate the H⁻ beam that at extraction radius crosses a stripper foil where the two electrons are removed. Some research laboratories are also building cyclotron which accelerates H⁻ at higher energies, and at TRIUMF since the 1975 H⁻ beams are accelerated up to 520 MeV. The main limit of cyclotrons which accelerate H⁻ is the maximum magnetic fields usable. Indeed a charged particle moving in a magnetic field B is subject to an electric field (or Lorenz force). The equivalent electric field in the rest frame of the H⁻ ion is given by

$$E = 3\beta\gamma B \text{ [MV/cm]} \quad (1)$$

Where $B$ is the static magnetic field in Tesla, $\beta=v/c$ and $\gamma=(1-\beta^2)^{-1/2}$ are the relativistic parameters of a particle with velocity v.
The probability to remove the second electron from the H⁻ or the lonely electron from the $H_2^+$ ion is given by the following formula [12]:

$$D = \frac{1}{2}\int_0^1 \exp(-\frac{\alpha}{\beta})d\mu \; ; \; \alpha = \frac{4}{3}\sqrt{2\frac{m}{h^2}W^{\frac{3}{2}}}/eE$$

Where: $\mu$ is the cosine of the angle between the electric field and the direction of the electron motion, $m$ and $e$ are the mass and charge of the electron, $W$ is the binding energy of the electron in the ion (H⁻ or $H_2^+$), and $E$ is the electric field due to the magnetic field (1).
To avoid the electromagnetic dissociation of H⁻, the magnetic field has to be lower and lower when the energy increases. So to accelerate H⁻ up to 1 GeV the maximum magnetic field acceptable is 3 kGauss and the radius of about 19 m. The binding energy of the electron of the $H_2^+$ molecule is about 20 time stronger than the H⁻, and consequently the use of magnetic field as high as 7 T even at energies as high as 1 GeV/amu is permitted. In Table 4 a comparison between the probability of electrical dissociation for H⁻ and $H_2^+$, for the commercial cyclotron and for the case of interest for DAEδALUS experiment are presented. The probability of electrical dissociation for commercial cyclotron was arbitrarily set to 1 and the probability for dissociation of $H_2^+$ accelerated at 800 MeV in an average magnetic field of 6T is 4 time less. We recall that the 30 MeV commercial cyclotrons are able today to supply a beam current up to 1.5 mA, so the amount of beam losses due to the electrical dissociation of $H_2^+$, with an average beam current of 5 mA, is lower than 1% the beam losses in the commercial cyclotron.

*Stripper mean life*

There are further important differences between the stripping process for H⁻ and $H_2^+$. In the H⁻ case both the two electrons have to be removed to extract one proton and the foil has to be thick enough to guarantee the stripping efficiency of 100% of all the particles which cross the stripper, while for $H_2^+$ if a molecule is not stripped at first cross through the stripper, it turns inside the cyclotron and hits once again the stripper until it is stripped. Then for $H_2^+$ it is possible to use a stripper with thickness thinner than for H⁻ and then a longer mean life is expected. According to the TRIUMF experience [13] a mean life higher than 10 hour is expected for an average beam current of 2 mA of $H_2^+$. This is a conservative limit. Indeed, the magnetic field to accelerate $H_2^+$ has reversed polarity compared to the field of H⁻, so the electrons stripped in the case of H⁻ are bent towards the centre of the machine and hit the stripper foil after spiralling in the magnetic field, while for $H_2^+$ the electrons are bent towards the outer radius. So if the orbit radius of the stripped electrons is larger than 4-5 mm, an electron catcher could be installed to remove these electrons and strongly reduce the stripper damage.
If we are able to place the stripper foil in a position where the field is lower than 1 T the beam radius of the electron are larger than 4 mm. According with the measured mean life of the stripper foils in the commercial cyclotron at 30 MeV, the electron damage of the stripper foils is the main source of stripper destruction. If we are able to stop the electron produced by the stripping process the mean life of the foil should be longer than the mean life measured in the small commercial cyclotron. Indeed at higher energy the energy lost by a beam particle crossing the foil is lower than at low energy.

Table 4: Comparison of parameters relevant to evaluate the electrical dissociation for H⁻ and $H_2^+$ beams

|  | H⁻ | $H_2^+$ |
| --- | --- | --- |
| Binding energy | 0.755 eV | 15.1 eV |
| Magnetic field | <1.3> T | <2.3> T |
| Energy (MeV/amu) | 30 | 800 |
| Electric field (MV/cm) | 0.998 | 10.8 |
| Dissociation probability | 1 | 0.0085 |

## BEAM LOSSES DUE TO RESIDUAL GAS

Due to the interactions with the residual gas, ions could loose the orbital electron along the acceleration path. The fraction of particles which survives is given by [14]:

$$T = N/N_0 = \exp(-3.35 \cdot 10^{16} \int \sigma_l(E) \, P \, dl)$$

$$\sigma_l(E) \approx 4\pi a_0^2 (v_0/v)^2 (Z_t^2 + Z_t)/Z_i$$

Where: P is the pressure (torr), L is the path length in cm. $\sigma_l(E)$ is the cross section of electron loss, $v_0$ and $a_0$ are the velocity and the radius of the orbit of Bohr respectively, and $Z_t$ and $Z_i$ are the atomic number of the residual gas and of the incident ion respectively. This formula is in good agreement with experimental data. Table 5 shows a comparison of the relevant parameters for the TRIUMF cyclotron and the Ring Superconducting Cyclotron (RSC) here described. The expected losses in percent should be 10 time less, while the proton current lost should be a factor 2 smaller. The expected beam power lost along the whole acceleration path should be lower than 4.3 kW. Despite this number seems a little high, we have to consider that the particle are lost along the acceleration path so there is not a specific hot point and that 3 kW of the power loss is transported by proton particles with energy higher than 400 MeV and with a range in iron longer than 150 mm, so the power it is released in the whole volume of the iron and not in a hot spot. Of course these beam losses could be reduced if a better vacuum and/or higher energy gain per turn are achieved. The proposed high energy cyclotron here discussed is more compact and smaller than the TRIUMF one, so to achieve a better vacuum seems feasible. Moreover a better operating vacuum is useful in order to increase the reliability of the RF cavities too.

Table 5: Beam losses due to interactions with residual gases, along the acceleration path.

|  | $E_{max}$ MeV | $\Delta E/\Delta n$ MeV | $R_{ex}$ m | <I> mA | Vac. torr | $I_{loss}$ % | $I_{loss}$ μA |
|---|---|---|---|---|---|---|---|
| TRIUMF | 520 | 0.34 | 7.8 | 0.4 | $2 \cdot 10^{-8}$ | 1.66 | 6.6 |
| RSC $H_2^+$ | 800 | 3 | 4 | 2 | $2 \cdot 10^{-8}$ | 0.15 | 3 |

## CONCLUSION

A lot of work has to be done to achieve the final design of an accelerator complex to deliver $H_2^+$ at 800 MeV/amu for the DAEδALUS experiment. The preliminary study here presented shows that it is a realistic goal. In particular the sector magnets able to produce an average magnetic fields of 2.3 T with the right magnetic field shape and the necessary coils to drive these sectors seem to be feasible even with the present technology, but the optimisation of the magnetic shape and size of the sectors to avoid the spiral shape and to reduce the magnetic forces on the superconducting coils must be completed. Significant reduction of both these problems could be achieved by increasing the extraction radius of about 10÷20%. Maybe this is the best solution which allows also for more room between the sectors where the RF cavities have to be installed. Of course the advantage to have straight sectors allows installing RF cavities similar to that built for the PSI which achieves a maximum voltage of 1 MV. The use of these cavities will reduce the number of turns in the ring cyclotron and the beam losses due the interaction with the residual gases.

Up to now the most critical point is the optimum vacuum required inside the acceleration chamber. Despite the good vacuum level achieved at TRIUMF cyclotron, the use of RF cavities to be operated at high voltage and high power, like the PSI ones, could limit this goal if not properly designed.

Cyclotrons for DAEδALUS experiment need to guarantee a high level of reliability, easy operation as well as high conversion efficiency from electrical to beam power.

We believe that extraction by stripping is a very powerful tool to increase the reliability and simplify the operation as demonstrated by the success of commercial cyclotrons.